\documentclass[twocolumn,english,pre]{article}
\usepackage{pslatex}
\usepackage[T1]{fontenc}
\usepackage[latin1]{inputenc}
\usepackage{graphicx}
\usepackage{amssymb}

 \newcommand{\lyxaddress}[1]{
   \par {\raggedright #1 
   \vspace{1.4em}
   \noindent\par}
 }
\usepackage{verbatim}

\usepackage{babel}

\begin{document}

\title{Tumors under periodic therapy -- Role of the immune response time
delay}

\author{D. Rodríguez-Pérez, Oscar Sotolongo-Grau, Ramón Espinosa Riquelme,
\\
Oscar Sotolongo-Costa, J. Antonio Santos Miranda,  J.C. Antoranz}

\date{~}

\maketitle

\lyxaddress{D. Rodríguez-Pérez $\cdot$ Oscar Sotolongo-Grau $\cdot$ Ramón Espinosa
Riquelme $\cdot$ J.C. Antoranz\\
Departamento de Física Matemática y Fluidos, UNED \\
POBox 60141, 28080 Madrid, Spain\\
Fax: 91-3987628 \\
E-mail: daniel@dfmf.uned.es\\
~\\
Oscar Sotolongo-Costa\\
Henri Poincaré Chair of Complex Systems, Havana University \\
Havana, Cuba\\
~\\
J. Antonio Santos Miranda\\
 Servicio de Oncología y Radioterapia, H.G.U. Gregorio Marañón\\
Madrid, Spain}

\begin{abstract}
We model the interaction between the immune system and tumor cells
including a time delay to simulate the time needed by the latter to
develop a chemical and cell mediated response to the presence of the
tumor. The results are compared with those of a previous paper, concluding
that the delay introduces new instabilities in the system leading
to an uncontrolable growth of the tumour. Then a cytokine based periodic
immunotherapy treatment is included in the model and the effects of
its dossage are studied for the case of a weak immune system and a
growing tumour. We find the existence of metastable states (that may
last for tens of years) induced by the treatment, and also of potentially
adverse effects of the dossage frequency on the stabilization of the
tumour. These two effects depend on the delay, the cytokine dose burden
and other parameters considered in the model.
\end{abstract}
\begin{description}
\item [Keywords]tumor growth, delay differential equations, immunotherapy,
immunodepression
\end{description}

\section{Introduction}

Immunotherapy is defined as the use of the immune system and its products
to prevent, treat and control some illness. Immunotherapy may act
stimulating the immune system of a patient with some kind of melanoma,
kidney cancer, etc\ldots{} in order to eliminate or control the population
of tumoral cells. 

The development of a cancer tumor under the influence of the immune
system deploys a very rich dynamics with some aspects that must be
highlighted: 

\begin{enumerate}
\item T-lymphocytes are the cells that perform the tumor elimination, but
to do that, another cells, the B- lymphocytes, must activate them
with the help of cytokines \cite{Tortora99,Dezfouli05,Saleh05}
\item When the tumor size increases, it induces a deactivation of lymphocytes
that enter the tumor region. This phenomenon is known as immunodepression
\item Obviously, in the whole process of interaction between the immune
system and the tumor, a delay between the detection of the {}``strange
cells'' (antigen) and the attack of the T-lymphocytes is present.
This delay will be considered as the control parameter in the system
under study \cite{Forys02,Forys03}. As we shall see, it is important
since it produces changes in the stability of the solutions of the
set of equations describing the system.
\end{enumerate}
In this paper, the model proposed in \cite{Sotolongo03} is modified
with the inclusion of a {}``memory effect'' in the term describing
the interaction between the immune system and the tumor. This leads
to formulate the problem in terms of a set of delay differential equations
\cite{Beretta02,Hale77,Gopalsamy92}. The influence of this memory
effect is studied both with and without therapy. Besides, the range
of variation of the biologic parameters is estimated, so that the
numeric simulations were performed inside the range.

As far as we know, the existence of a delay between the stimulus (antigen)
and the triggering of the defenses of the immune system has not been
considered enough in the literature, despite the excellent reports
in \cite{Galach03,Byrne97}, where the effect of time delay in the
dynamics of tumor- immune system is included in the model proposed
in \cite{Kuznetsov94}. There, it is shown that time delay is a very
important factor to take into account in the modelling of the immune
system and the reaction of living organisms to diseases. Another delay
times have been shown to be of importance. For example, in \cite{Villasana03}
this characteristic time is introduced in a model adapted from \cite{Kuznetsov94,Kirschner98}
to take into account the cell cycle, leading to a linear delayed term
in the equations. In the same way, the study presented in the this
paper is based on the model introduced by some of the authors in \cite{Sotolongo03}
to describe the effects of tumor immunodepression, now including a
time characterising the immune system delayed response to the tumor.

\section{The model}

\subsection{Construction of the model }

In \cite{Sotolongo03}, a Lotka-Volterra model with the inclusion
of some additional terms was proposed as: 

\begin{equation}
\begin{array}{rl}
\frac{dX}{dt}= & aX-bXY\\
\frac{dY}{dt}= & dXY-fY-kX+u+F\cos^{2}wt\end{array}\label{eq:Lotka-Volterra-depress-forced}\end{equation}
where $X(t)$, and $Y(t)$ are the populations of tumor cells and
T-lymphocytes, respectively. The growing rate of tumor cells is proportional
to $X$ and the mortality rate is proportional to the frequency of
interaction with the lymphocytes. The growing rate of lymphocytes
is proportional to the interaction with the tumor cells and to the
flux rate ($u$) of lymphocytes to the region where the tumor is localized.
Their death rate is linked in this model to two terms: natural death
($-fY$) and the increase of the tumor mass, which may induce a depression
of the immunity system ($-kX$). A characteristic of this system is
that the immunodepression has been taken into account. Immunodepression
may lead to fatal consequences since it may annihilate the whole immune
system, mathematically expressed as $Y(t)=0$. (In this model, the
population of lymphocytes may even become negative after being null,
so that in that case it makes no sense to follow the integration).
To introduce the effects produced by the periodic therapy with cytokines
we choose, as in \cite{Sotolongo03}, the term $F\cos^{2}wt$, as
a first model of a periodic positive function, being $w$ the frequency
of the therapy and $F$ the peak value of the dose.

The term $dXY$ is specially important for this study. It mimics the
interaction between the two populations, $X$ and $Y$, with a frequency
$d$ of recognition of malignant cells by the immune system. We consider
the effect of time delay for this chemical signal mediated interaction
which, according to \cite{Galach03}, introduces here the values $X(t-T)$
and $Y(t-T)$. The evolution equations (\ref{eq:Lotka-Volterra-depress-forced})
become now: \begin{equation}
\begin{array}{rl}
\frac{dX}{dt}= & aX-bXY\\
\frac{dY}{dt}= & dX(t-T)Y(t-T)-fY-kX+u+G(t)\end{array}\label{eq:Lotka-Volterra-depress-delay-forced}\end{equation}
where $G(t)=F\cos^{2}wt$, and $T$ is the time the immune system
takes to react over the malignant cells once recognized. We want to
point out that the delay time is absent in the other interaction term
$bXY$, since it represents the direct action of the already present
T-lymphocytes and, therefore, it already exist at time $t$.

If $F=k=u=0$, we recover the classical delayed Lotka-Volterra model
\cite{Driver77} with characteristic time $t_{c}=1/\sqrt{af}$. This
time will be used to rescale the system (\ref{eq:Lotka-Volterra-depress-delay-forced})
as:\begin{equation}
\begin{array}{rl}
\frac{dx}{d\tilde{t}}= & \alpha x-xy\\
\frac{dy}{d\tilde{t}}= & x(\tilde{t}-\tau)y(\tilde{t}-\tau)-\frac{1}{\alpha}y-\kappa x+\sigma+H(\tilde{t})\end{array}\label{eq:dimless-system}\end{equation}
being\[
H(\tilde{t})=V\cos^{2}\beta\tilde{t}\]
where\[
\tilde{t}=t/t_{c},\qquad\tau=T/t_{c},\qquad x=\frac{d}{\sqrt{af}}X,\qquad y=\frac{b}{\sqrt{af}}Y\]
and\[
\kappa=\frac{kb}{d\sqrt{af}},\qquad\alpha=\sqrt{\frac{a}{f}},\qquad\sigma=\frac{ub}{af}\]
\[
V=\frac{Fb}{af},\qquad\beta=\frac{w}{\sqrt{af}}\]
To simplify notation, from now on we make $\tilde{t}=t$.

\subsection{\label{sub:parameter-estimation}Estimation of some parameters}

To get some insight about the numerical range of the parameters to
include in the model, let us estimate the values of the parameters
$a$, $f$, $T$, $u$, $b$, $d$, $k$ as well as $X_{0}$, using
days as time unit.

It is known that the time a solid tumor takes to reach twice its initial
volume is about $70\textrm{ days}$ \cite{Begg77}. Obviously, this
{}``representative'' value is useful only to have an idea of the
orders of magnitude; there are tumors that duplicate their volume
in $20\textrm{ days}$, whereas others can do it in $100\textrm{ days}$.
Tumors like those of colon or rectal cancer take a few years to duplicate.
We admit that in the absence of lymphocytes the growth is exponential
so our parameter $a=\log2/T_{d}$ $(\textrm{days})^{-1}$ will be
bounded to the interval $a\in[10^{-4},\,10^{-2}]\textrm{ days}^{-1}$.
This is a conservative estimation, compared to others found in the
literature \cite{dePillis05}. Experimentally, the values for the
lifetime of lymphocytes can be measured, being about $30\textrm{ days}$
\cite{Kuznetsov94}. Then it can be assumed that $f\in[10^{-2},\,10^{-1}]\textrm{ days}^{-1}$.
The time the lymphocytes take to proliferate is between $2$ and $12\textrm{ days}$
\cite{Byrne97}. This time can be taken as the delay time $T$ of
the immune system. Assuming that, in the absence of tumor, there is
an already established stationary state, the second equation in (\ref{eq:Lotka-Volterra-depress-delay-forced})
shows that in the initial state: $u-fY=0$. It is known that in the
spleen of a laboratory rat there are approximately $10^{8}$ lymphocytes,
but only part of them are cytotoxic T-lymphocytes. Let us cautiously
evaluate this fraction as about $3\%$ \cite{Forys02,Sotolongo03,Galach03},
so that $Y_{0}\simeq3\cdot10^{5}\textrm{ cells}$. With these data
it can be estimated that $u\simeq1.2\cdot10^{4}\textrm{ cells}\cdot\textrm{day}^{-1}$.
As in \cite{Kuznetsov94,Kirschner98}, other parameters can be guessed
so that $b\in[10^{-9},\,10^{-7}]\textrm{ cells}^{-1}$, and $d$ is
then of the same order as $b$ and with the same dimensions. Besides,
considering that $k$ represents the fraction of tumor cells that
block the flow of lymphocytes to the region occupied by the tumor,
it can be assumed the existence of a initial population $X_{0}$ of
about $10^{6}\textrm{ cells}$ and, in the absence of lymphocytes,
we can make an initial appreciation of $k\simeq1.2\cdot10^{-2}\textrm{ days}$.
All these estimates are in good agreement with particular values collected
in the literature (see \cite{dePillis05} and references therein).

These estimates of the parameters in (\ref{eq:dimless-system}), are
shown in table \ref{table:dimless-parameters}.

\begin{center}%
\begin{table}[htb]
\begin{center}\begin{tabular}{|c|c|c|c|}
\hline 
&
min&
max&
typical\\
\hline
\hline 
$\alpha$&
$\sim10^{-2}$&
$\sim1$&
$1.5$\\
\hline 
$\kappa$&
$\sim10^{-2}$&
$\sim10$&
$0.95$\\
\hline 
$\sigma$&
$\sim10^{-2}$&
$\sim10^{3}$&
$0.5$\\
\hline 
$\tau$&
$\sim10^{-3}$&
$\sim10^{2}$&
$\sim1$\\
\hline 
$x_{0}$&
$\sim10^{-3}$&
$\sim10^{2}$&
$0.1$\\
\hline
\end{tabular}\end{center}

\caption{\label{table:dimless-parameters}Ranges for the orders of magnitude
of the dimensionless parameter values of equation (\ref{eq:dimless-system})
as well as initial condition for tumour cell population. In the last
column, typical values used in the simulations reported in this work.}
\end{table}
\end{center}

\section{Results }

\subsection{System with $G(t)=0$ and without delay (from \cite{Sotolongo03})}

Our system (\ref{eq:dimless-system}) without delay and with $G(t)=0$
becomes:\begin{equation}
\begin{array}{rl}
\frac{dx}{dt}= & \alpha x-xy\\
\frac{dy}{dt}= & xy-\frac{1}{\alpha}y-\kappa x+\sigma\end{array}\label{eq:dimless-system-unforced-undelayed}\end{equation}
The stability of the stationary states of this system can be expressed
as:

\begin{enumerate}
\item State $L_{0}=(0,\,\alpha\sigma)$, tumorless state\\
If $\sigma>1\;\Rightarrow$ Stable node \\
If $\sigma<1\;\Rightarrow$ Saddle point \hfill(E1)
\item State $L_{1}=\left(\frac{1-\sigma}{\alpha-\kappa},\,\alpha\right)$\\
If $\frac{\kappa}{\alpha}>\sigma>1$ or $\sigma>\frac{\kappa}{\alpha}>1\;\Rightarrow$Saddle
point \hfill(E2) \\
If $\frac{\kappa}{\alpha}<\sigma<1\;\Rightarrow$ Stable node or focus
\hfill(E3) \\
If $\sigma<\frac{\kappa}{\alpha}<1\;\Rightarrow$Unstable node or
focus \hfill(E4) 
\end{enumerate}

\subsection{Delayed system without treatment ($G(t)=0$)}

In this case our system becomes: \begin{equation}
\begin{array}{rl}
\frac{dx}{dt}= & \alpha x-xy\\
\frac{dy}{dt}= & x(t-\tau)y(t-\tau)-\frac{1}{\alpha}y-\kappa x+\sigma\end{array}\label{eq:dimless-system-unforced}\end{equation}
Current techniques of delay differential equations give the following
analytic results concerning stability of solutions \cite{Beretta02,Hale77,Gopalsamy92}: 

\begin{enumerate}
\item The stability of the tumorless state $L_{0}$ is unaltered. Then (E1)
is fulfilled $\forall\,\tau\ge0$
\item The state $L_{1}$ is kept unstable $\forall\,\tau\ge0$ in all cases
given by (E2).
\item In the stability range given by (E3), a value of $\tau$, \emph{i.e.}
$\tau_{c}$, exists such that the stationary state $L_{1}$ becomes
unstable.
\item For the range given by (E4), the unstable state $L_{1}$ preserves
its instability.
\end{enumerate}
\begin{figure}[htb]
\begin{center}\includegraphics[%
  width=0.90\columnwidth,
  height=0.55\columnwidth]{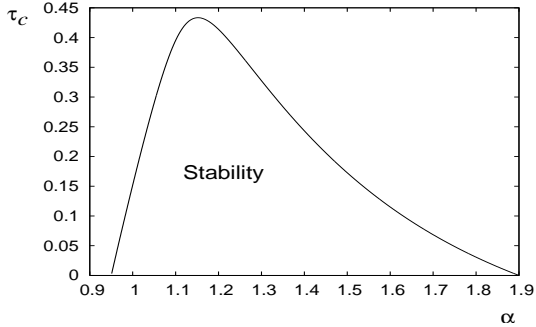}\end{center}

\caption{\label{cap:tau-alpha-stability}Function $\tau_{c}(\alpha)$, the
threshold above which the system (\ref{eq:dimless-system-unforced})
becomes unstable. Here the values of $\kappa=0.95$, $\sigma=0.5$,
within the physiological estimated range, were taken such that the
system is stable for $\tau=0$. No immunological treatment ($V=0$)
was considered.}
\end{figure}

The value of $\tau=\tau_{c}$ for which $L_{1}$ becomes destabilized
due to the presence of the delay can be computed analytically (see
\cite{Beretta02} for details, or the appendix in \cite{Villasana03}).
Figure \ref{cap:tau-alpha-stability} shows these values of $\tau$
for which the system becomes unstable, as a function of the parameter
$\alpha$ (the same can be done applying the criteria in \cite{Beretta02}
for the other parameters $\kappa$ and $\sigma$). These values of
$\tau_{c}(\alpha)$ can be computed analytically within the range
of parameters for which $L_{1}$ has positive abscissa and where $L_{1}$
is stable for $\tau=0$.

\subsection{\label{sub:Results-with-delay}Results with delay and therapy}

To make a useful phase diagram, numerical integration of the system
(\ref{eq:dimless-system}) can be performed for different values of
the parameters and explore the values of $\beta$ and $\tau$ for
which, given a set of values of the other parameters, the system becomes
stable or unstable.

As the initial condition for the integration of the system of delay
differential equations, the steps method can be used for which initial
conditions need to be specified, giving the values of $x(t)$ and
$y(t)$ in the interval $[-\tau,\,0)$. We will take as a realistic
initial condition, $x(t)=x_{0}(0)e^{\alpha(1-\sigma)t}$, $y(t)=\alpha\sigma$,
in $t\in[-\tau,\,0)$, corresponding to an exponentially growing tumour,
and an unaware immune system ($\tau$-delayed response).

We will perform the numerical integration of the system up to a time
$t_{\mathrm{max}}$ corresponding to $\sim25\textrm{ years}$ (from
typical values estimated in \ref{sub:parameter-estimation}, it can
be obtained $t_{\mathrm{max}}\in[30,\,10^{3}]$; we will consider
$t_{\mathrm{max}}\leq100$ in what follows) . However, whenever the
condition $y\leq0$ is met, that is, when the immune system becomes
anihilated due to tumor agressiveness, integration will stop before
$t=t_{\mathrm{max}}$.

\begin{figure}[htb]
\begin{center}\includegraphics[%
  width=0.90\columnwidth,
  height=0.55\columnwidth]{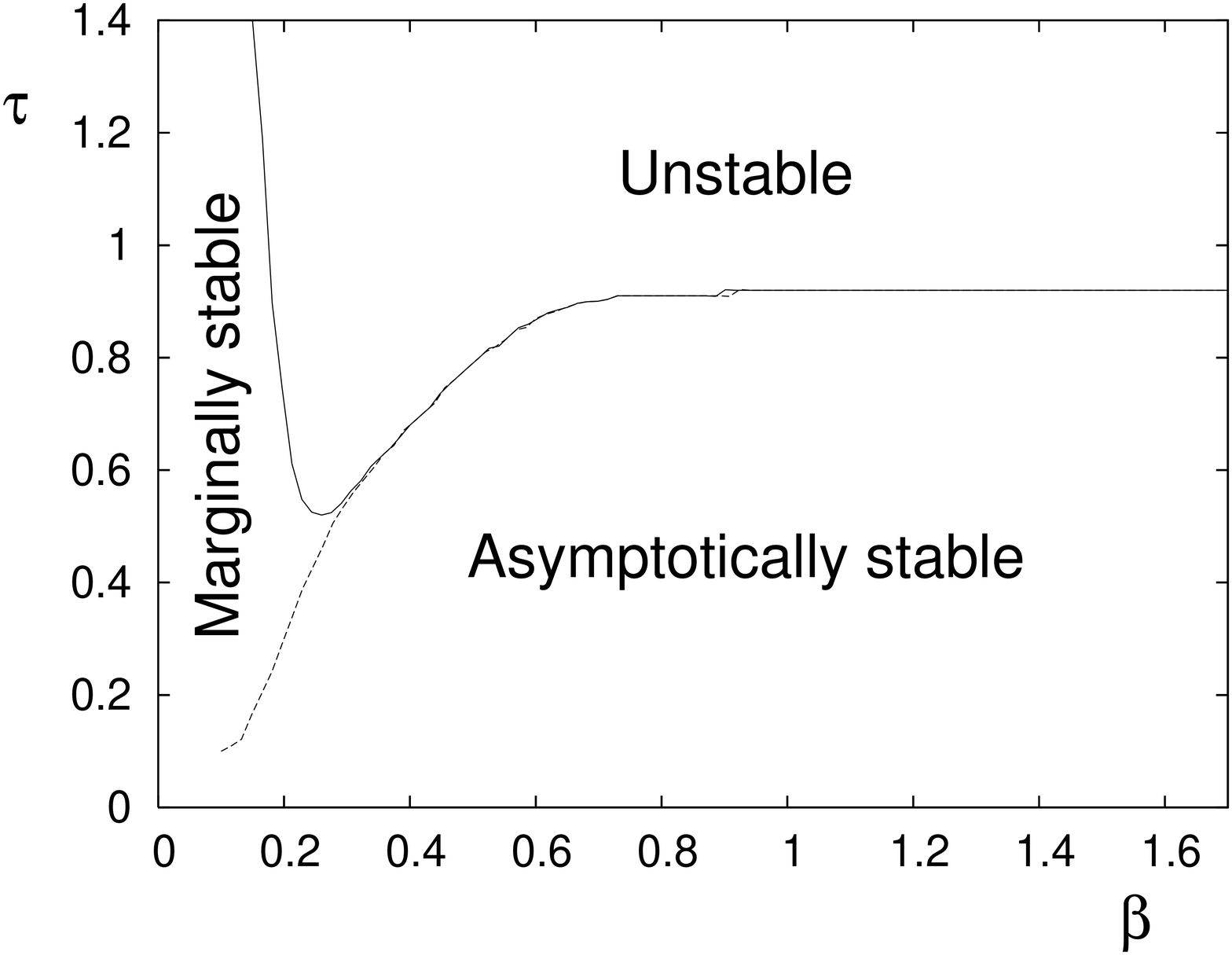}\end{center}

\caption{\label{cap:beta-tau-stability}Representation of stable and unstable
regions in the phase plane $(\beta,\,\tau)$. The values of the other
parameters in this case are: $\kappa=0.95$, $\sigma=0.5$, $\alpha=1.5$,
$V=0.5$, all of them within the estimated physiological range. The
initial conditions were thus taken as $x_{0}=0.1$, $y_{0}=\sigma\alpha=0.75$.
Dashed line delimits the asymptotically stable region where tumour
size remains controlled after $t=100$. Solid line limits the {}``marginally
stable'' region, where tumor grows in an uncontrollable way after
a dimensionless time $t_{\mathrm{max}}=10$.}
\end{figure}

The result of the integration of (\ref{eq:dimless-system}) with these
initial conditions and this stop conditions ($y>0$ and $t<t_{\mathrm{max}}$),
quite surprisingly, leads to a division of the $(\beta,\,\tau)$ plane
into three regions instead of two as could be expected, with an interesting
{}``transition region'' in which, being the system unstable, the
instability manifests itself only {}``asymptotically'', \emph{i.e.},
only when a very long time has elapsed.

The region marked as stable in figure \ref{cap:beta-tau-stability}
denotes that the system remains stable at any {}``biologically meaningful''
time, that is, tumour growth is controlled by the cytokine treatment.
The one marked as unstable denotes, for the chosen values of the parameters,
those values of the frequency $\beta$ and the delay $\tau$ for which,
the system becomes unstable in a finite time (less than $t_{\mathrm{max}}$)
leading to an unlimited tumor growth and total annihilation of the
immune system \cite{Sotolongo03}.

The marginally stable region corresponds to unstable states where
instabilities manifest themselves after such a long time (for $t>t_{\mathrm{max}}$)
that they have not a practical sense. In the above case, inside this
region the instabilities manifest only for times in the order of tens
of years, an unpractical value in the medical sense. The limit between
this marginally stable region and the unstable one in figure \ref{cap:beta-tau-stability},
depends strongly on the initial conditions ($x_{0}$, $y_{0}$ as
well as on the hipotheses about their history) used for the numerical
calculation.

\section{Discussion and Conclusions}

We have studied the effect of the chemical signal and B-lymphocytes
mediated interaction between (T-)lympho\-cytes and cancer cells (that
is, the time delay) upon the development and growth of a cancer tumor.
We have found that an increase in this body reaction time will unstabilize
a system in which the tumor has evolved to a latent state. As far
as this delay is a body characteristic time, it will classify cancer
situations into stable (that is, limited by the immune system and
thus curable) and unstable (those which cannot be controlled by the
immune system). These states will be characterized by the particular
dynamical system parameter values, which must be estimated as shown
in section \ref{sub:parameter-estimation} from real available data.
Furthermore, the mathematical form of the evolution equations determines
parameter combinations (in the form of dimensionless groups) which
are more suitable to describe and classify the dynamics.

For a fixed time delay, numerical simulations show that tumors evolve
under cytokine treatment as follows. For an initially unstable system,
above the $\tau_{c}$ critical value, immunotherapy treatment helps
to stabilize this state. For low dossage frequency $\beta$, tumor
evolves so slowly that the system behaves as stable for times larger
than a given $t_{\mathrm{max}}$ (taken of the order of decades, see
conditions in \ref{sub:Results-with-delay}). Above a threshold the
system becomes also asymptotically stable. Therapy does not lead to
the full anihilation of tumor cells, but keeps tumor size in a controlled
state, allowing the inclusion of other methods, like surgery, or more
directed therapies like radiation, aimed to a localized and controlled
tumoral mass. 

Finally, when the system is not stable but instability does not lead
to uncontrolled tumor growth for, at least, $t_{\mathrm{max}}$ (taken
of the order of two decades), $\beta$ dossage values larger than
a critical one may lead to system unstabilization as well as to its
stabilization if the frequency is further increased, in a way sensitive
to the initial state of the tumour and the particular value of $\tau$.
In general, we conclude that there exist one critical low dossage
frecuency that makes the system unstable (for limited times smaller
than a $t_{\mathrm{max}}$) and another higher frequency that takes
the system back to stability (long time stability, in this case).
These particular values will depend on the immune system delay ($\tau$),
as well as on the dose burden ($V$) and other parameters of the system. 

It is of paramount importance to notice that for some practical range
of citokine dossage frequencies ($\beta\gtrsim0.75$ in the particular
case shown in figure \ref{cap:beta-tau-stability}), a particular
value of the immune system delay may fall above the region of stability
(asymptotic as well as transitory), making the use of immunotherapy
useless. 

~

\section*{Acknowledgements}

This study was supported by research grants FIS-G03/185, FIS-04/1885
and FIS-PI1192/2005 of the \emph{Fondo de Investigación Sanitaria},
Instituto Carlos III, Madrid, Spain.

\end{document}